\documentclass[sigconf]{acmart}

\usepackage{xcolor}
\newcommand{\new}[1]{\textcolor{black}{#1}}
\AtBeginDocument{%
  \providecommand\BibTeX{{%
    \normalfont B\kern-0.5em{\scshape i\kern-0.25em b}\kern-0.8em\TeX}}}

\acmConference[]{}
\acmBooktitle{}
\acmPrice{}
\acmISBN{}

\begin{document}

\title{Hydrogel-based Bio-nanomachine Transmitters for Bacterial Molecular Communications}

\author{Daniel P. Martins}
\affiliation{%
  \institution{\new{VistaMilk SFI Research Centre\\ Telecommunications Software and Systems Group}}
  \city{Waterford}
  \country{Ireland}}
 \email{dpmartins@tssg.org}

\author{Huong Q.-O'Reilly}
\affiliation{%
  \institution{\new{Telecommunications Software and Systems Group \\Pharmaceutical and Molecular Biology Research Centre}}
  \city{Waterford}
  \country{Ireland}}
\email{horeilly@tssg.org}

\author{Lee Coffey}
\affiliation{%
  \institution{\new{VistaMilk SFI Research Centre\\ Pharmaceutical and Molecular Biology Research Centre}}
  \city{Waterford}
  \country{Ireland}}
\email{lcoffey@wit.ie}

\author{Paul D. Cotter}
\affiliation{%
  \institution{\new{VistaMilk SFI Research Centre\\ Teagasc Food Research}}
  \city{Fermoy}
  \country{Ireland}}
\email{paul.cotter@teagasc.ie}

\author{Sasitharan Balasubramaniam}
\affiliation{%
  \institution{\new{VistaMilk SFI Research Centre\\ Telecommunications Software and Systems Group}}
  \city{Waterford}
  \country{Ireland}}
 \email{sasib@tssg.org}


\begin{abstract}
Bacterial quorum sensing can be engineered with a view to the design of biotechnological applications based on their intrinsic role as a means of communication. We propose the creation of a positive feedback loop that will promote the emission of a superfolded green fluorescence protein from a bacterial population that will flow through hydrogel, which is used to encapsulate the cells. These engineered cells are heretofore referred to as bio-nanomachine transmitters and we show that for lower values of diffusion coefficient, a higher molecular output signal power can be produced, which supports the use of engineered bacteria contained within hydrogels for molecular communications systems. In addition, our wet lab results show the propagation of the molecular output signal, proving the feasibility of engineering a positive feedback loop to create a bio-nanomachine transmitter that can be used for biosensing applications.
\end{abstract}



\maketitle

\section{Introduction}

Bacteria \new{utilise} signalling mechanisms (\new{i.e.,} quorum sensing) to drive collective behaviours within \new{homogeneous and heterogeneous} microbial populations. These biological communications systems have been investigated for the past 50 years, and recently they became the focus of biotechnological solutions design for  biofabrication and biosensing \cite{Liu2012,Holowko2016,Lentini2014,Luo2012,Raut2015}. For example, soft matter (i.e. gelatin) was assembled using enzymes emitted by bacterial cells, and \textit{Escherichia coli} bacteria were engineered with \new{a signalling system} used by \textit{Vibrio cholerae} for a timely detection of cholerae infection \cite{Liu2012,Holowko2016}. These \new{and other} examples highlight the range of applications that can be designed and built based on the engineering of the bacterial signalling mechanisms. 

Bacteria signalling \new{has also been} investigated using communications theory concepts, in a paradigm named as Molecular Communications \cite{Cobo2010,Akan2016,Unluturk2015,Einolghozati2016,Guo2016,Abadal2011,Grebenstein2018}. \new{From} this perspective, the bacteria are considered as biological bio-nanomachines that are engineered to process, emit and detect specific molecules, acting as transceivers found in conventional communication systems \cite{Cobo2010,Akan2016,Abadal2011,Guo2016}. For instance, \new{bacterial quorum sensing has} been used to create logic circuits, modulators and network links between bacterial nodes \cite{Martins2019,Unluturk2015,Einolghozati2016,Grebenstein2018}. Here our focus is on the processing and emission of molecular signals \new{as we} propose the design of a bio-nanomachine transmitter that can enable the development of safer intrabody molecular communications systems in humans and animals. Therefore, we investigate a physical model of a bio-nanomachine transmitter embedded in a hydrogel bubble, which will protect the host environment from mixing with engineered cells, and analyse its suitability for molecular communication systems. 

\begin{figure}[t!]
  \includegraphics[width=\columnwidth]{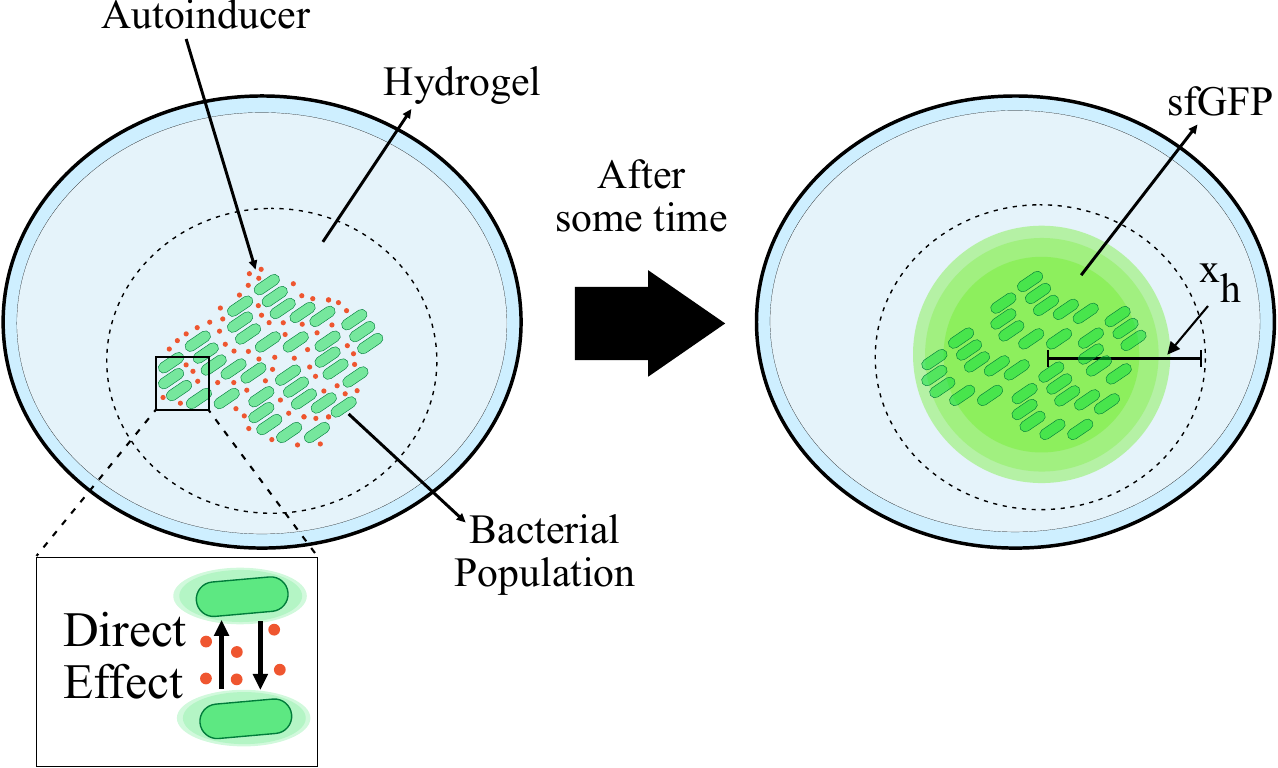}
  \caption{\new{The bio-nanomachine transmitter are embedded in hydrogel to protect them. The cells are engineered to directly exchange signals and produce sfGFP for minutes or hours depending on the signalling system used and the quantity of the cells.}}
  \label{fig:biotrans}
\end{figure}

This work is inspired by our previous work where we proposed the production and emission of molecular signals to attract bacteria to a specific location \cite{Martins2016}. In this paper, we focus our investigation on the production and emission of \new{a} specific molecular signal through the hydrogels. A visual representation of the proposed work can be seen in Figure \ref{fig:biotrans}, where a bacterial population is placed in a plate with nutrients to \new{grow} and produce the desired molecular output signal, a fluorescence protein (superfolder green flourescence protein--sfGFP). The performance of such \new{a} bio-nanomachine transmitter is evaluated \new{with} respect to the molecular signal throughput that is able to reach the border of the hydrogel bubble, and we also study the bacterial growth and nutrient consumption associated with the emission of molecules. The contributions of this paper include:
\begin{itemize}
    \item {\bf Bio-nanomachine transmitter physical model:} We investigate the molecular signals, the processes related with the emission of molecular signals, its throughput and signal power depending on the viscosity of the hydrogel bubble and internal chemical reactions.
    \item {\bf Proof-of-concept design:} A wet lab experiment is devised to demonstrate the emission of a molecular signal, sfGFP, from the bacterial population through the hydrogel. 
\end{itemize}

This paper is organised as follows. In Section \ref{sec:bio_design}, we characterise and review the state-of-the-art of biological transmitters. The communications and biological models are introduced in Section \ref{sec:sys_model}. Section \ref{sec:num_analysis} presents the numerical analysis of the molecular emission and throughput. A proof-of-concept design and analysis are presented in Section \ref{sec:exp_analysis}. Lastly, Section \ref{sec:conclusions} presents our conclusions.

\section{Bio-nanomachine Transmitter Design}\label{sec:bio_design}

Bacteria produces and consumes different molecular signals through chemical reactions to collectively establish, protect and maintain a community, either single- or multi-species. The internal machinery that enable these processes can be engineered to produce specific molecules that will be emitted and propagated through the chosen medium towards other bacterial populations \cite{Mehta2009,Smanski2016}. To coordinate these processes, bacteria often utilises an internal signalling mechanism know as quorum sensing, which is an exchange of molecular signals that drive bacteria towards the execution of collective behaviours, such as virulence factors production and biofilm formation \cite{Bassler1999,Martins2016,Martins2018}. The engineering of this process is the target for the design and construction of a bio-nanomachine transmitter. 

Bacterial populations grow and produce quorum sensing molecules by consuming the nutrients available in its vicinity. Therefore, a faster nutrient consumption will result in a faster population growth and higher production of quorum sensing molecules\new{. However}, this can affect the sustainability of the bacterial population due to the depletion of its energy source \cite{Wang2019,Martins2016,Hornung2018}. To avoid such situations, bacterial \new{populations coordinate their} collective behaviour depending on this resource. This coordination is described through a set of chemical reactions that involves strain-specific quorum sensing molecules and receptors, such as LuxI and LuxR (luciferase inducer and regulator, respectively) \cite{Bassler1999}. By engineering this machinery, we are able to create a positive feedback loop using the bacteria quorum sensing process to emit a molecular output signal, sfGFP, and the bacterial population will act as a bio-nanomachine transmitter. Figure \ref{fig:comms_model} shows a representation of this process, where bacteria produces LuxI and detects its concentration, using LuxR, creating the positive feedback loop. After being produced through this process, the molecular output signal, sfGFP, is then emitted from the bacterial population towards the membrane edge of the hydrogel bubble. 

The engineering of quorum sensing systems through feedback loops have been proposed before. In \cite{Tu2008}, a negative feedback loops have been designed to control the quorum sensing dynamics of \textit{Vibrio harveyi} bacteria. Moreover, a positive feedback loop was engineered to built a whole cell biosensor to detect mercury levels \cite{Cai2018}.  \new{Here, we propose the design of} a positive feedback loop to characterise a bacterial population as a bio-nanomachine transmitter. One important aspect of our design is the encapsulation of the bacterial population in a hydrogel bubble, which is a medium with a higher viscosity value and have been previously applied to protect bacterial populations  \cite{Li2017,Seliktar2012}. Hydrogels are polymers with infiltrated water that can be applied to encapsulate living cells and have permeability for a wide range of molecules and facilitate the design of contained quorum sensing applications \cite{Li2017,Seliktar2012}. For our \new{system, we propose the use of hydrogel} to create an environment where the bacterial population \new{undergoes} controlled growth and emits molecular signals, and, at the same time, they are protected against external attacks.

\section{System Model}\label{sec:sys_model}

\begin{figure}
  \includegraphics[width=0.8\columnwidth]{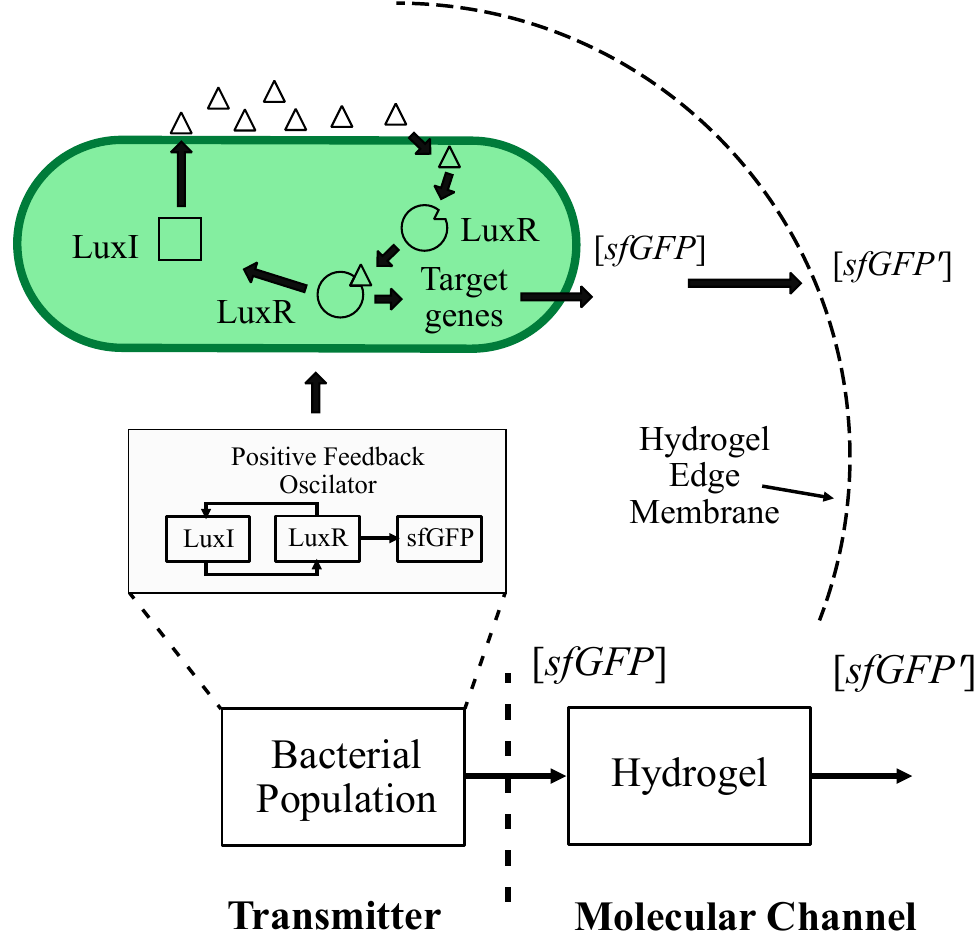}
  \caption{Representation of the molecular communications system used to model the emission of molecular signals by the bacterial population. Adapted from \cite{Bassler1999}}
  \label{fig:comms_model}
\end{figure}

As presented in Section \ref{sec:bio_design}, bacteria produce molecular signals by a series of chemical reactions using their internal machinery, which can be engineered for that purpose. In this paper, we engineer the \new{\textit{E. coli}} population internal machinery using a positive feedback loop to exploit their molecular communications and produce the desired molecular output signal, the fluorescence protein sfGFP. Here we describe the positive feedback loop using a set of chemical reactions (\ref{eqn:dA})-(\ref{eqn:dN_1}) and the propagation of the molecular output signal by solving the two-dimensional Fick's law of diffusion to characterise this molecular communications system. We describe the positive feedback loop using LuxI and LuxR, represented in Figure \ref{fig:comms_model}, as follows \cite{Melke2010}

\begin{equation}\label{eqn:dA}
\dfrac{d[A]}{dt} = c_A +\dfrac{k_A[C]}{K_A+[C]}- k_0[A]-k_1[R][A]+k_2[RA]
\end{equation}
\begin{equation}\label{eqn:dR}
\dfrac{d[R]}{dt} = c_R +\dfrac{k_R[C]}{K_R+[C]}- k_3[A]-k_1[R][A]+k_2[RA]
\end{equation}
\begin{equation}\label{eqn:dRA}
\dfrac{d[RA]}{dt} =  k_1[R][A]-k_2[RA]-2k_4[RA]^2+2k_5[C]
\end{equation}
\begin{equation}\label{eqn:dC}
\dfrac{d[C]}{dt} =k_4[RA]^2+k_5[C]
\end{equation}
where $[A]$, $[R]$, $[RA]$, $[C]$ are the \textit{AHL}, \textit{LuxR}, \textit{LuxR~--~AHL} complex and dimerized complex concentration, respectively; $c_A$ and $c_R$ are the transcription basal levels for \textit{AHL} and \textit{LuxR}, respectively; $k_A$ and $k_R$ are the transcription rates; $K_A$ and $K_R$ are the degradation rates; $k_0, k_1, k_2,k_3, k_4, k_5$ are the translation rates. The rate of molecular signal production, $[sfGFP]$, by the bacterial population that results from the quorum sensing process is represented as 
\begin{equation}\label{eqn:A1}
\dfrac{d[sfGFP]}{dt} =k_{sfGFP}\dfrac{[C]}{[C]+K_C}.
\end{equation}
In addition to the production of the sfGFP, the bacterial population will also consume nutrients to grow, and the rate of this process is represented by
\begin{equation}\label{eqn:dG1} 
\dfrac{d[GC]}{dt} = \left(\mu_1\dfrac{[N]}{K_{N}+[N]}-m_1\right)[GC].
\end{equation}
Then, the nutrient consumption rate can be evaluated as follows
\begin{equation} 
\begin{split}\label{eqn:dN_1}
\dfrac{d[N]}{dt} &= -U_{G1}\left(\mu_1\dfrac{[N]}{K_{N}+[N]}\right)[GC]-U_{AHL}\dfrac{d[A]}{dt}\\[2mm]
&-U_{LuxR}\dfrac{d[R]}{dt}-U_{sfGFP}\dfrac{d[sfGFP]}{dt},
\end{split}
\end{equation}
where $U_{G1}$, $U_{AHl}$, $U_{LuxR}$, and $U_{sfGFP}$ are utility parameters that represent the nutrient cost of the population growth, autoinducer, receptor and molecular output signal production, respectively. After reaching a high concentration, the molecular output signal $[sfGFP]$ propagates to the membrane edge of the hydrogel, from where it can freely diffuse to other engineered or natural bacterial cells. The propagation through the hydrogel can be modelled as \cite{Llatser2014}.
\begin{equation}\label{eqn:diffusion}
[sfGFP']=\dfrac{[sfGFP]}{\sqrt{4\pi D_ht_h}}e^{\dfrac{-x_h^2}{4D_ht_h}},
\end{equation}
where, $D_h$, $t_h$, and $x_h$ are the diffusion coefficient, the duration and the distance travelled by the molecular signals $[sfGFP]$ in the hydrogel $h$ channel, respectively. 

\begin{figure}
  \includegraphics[width=\columnwidth]{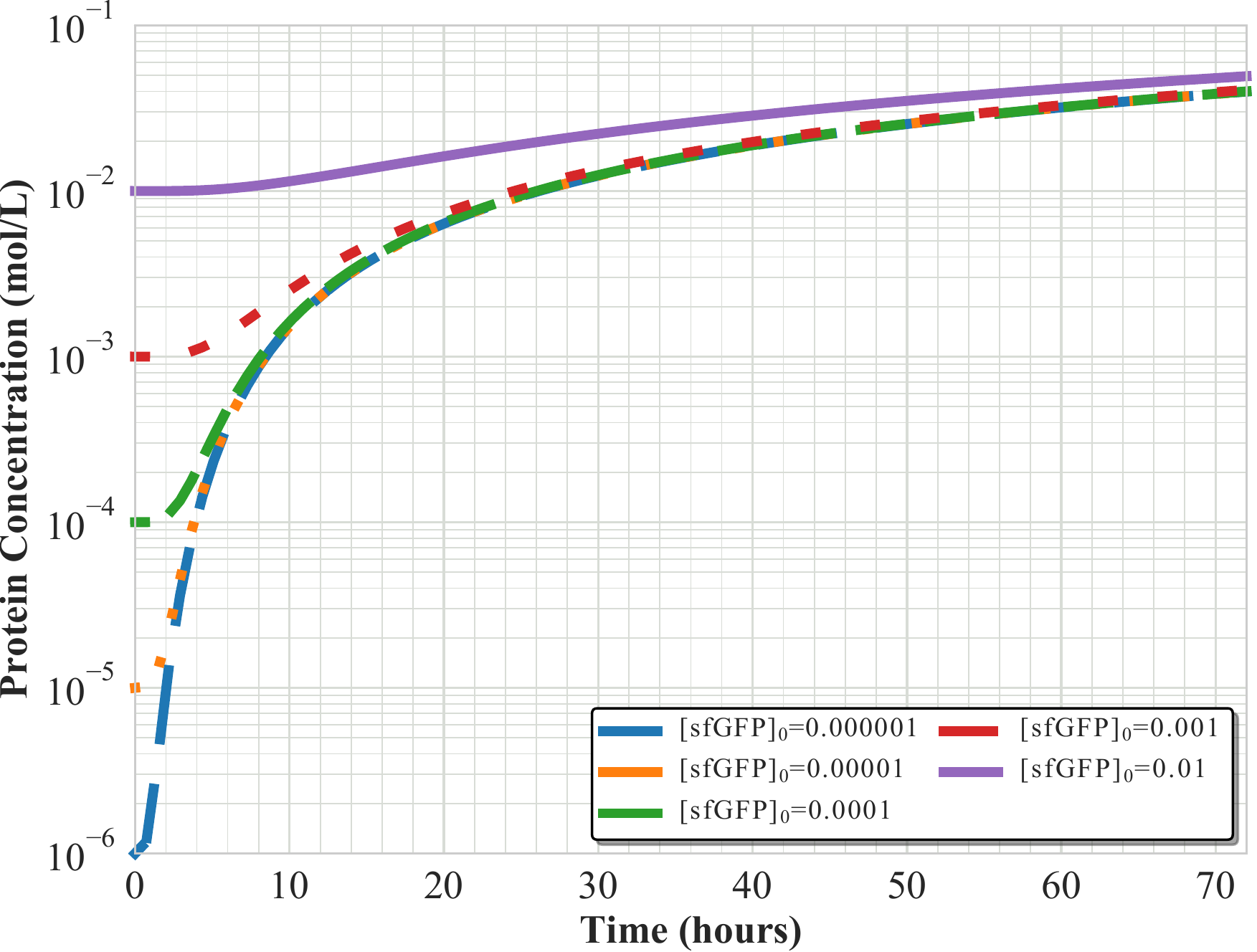}
  \caption{Evaluation of the molecular output signal throughput for a range of initial concentration values of $[sfGFP]$}
  \label{fig:throughput}
\end{figure}

\section{Numerical Analysis}\label{sec:num_analysis}

We evaluate the performance of this communications system using two metrics: the sfGFP throughput from the bacterial population and the signal power that reaches the hydrogel edge membrane. The throughput is evaluated directly from (\ref{eqn:dG1}), while the signal power is evaluated using (\ref{eqn:diffusion}) and converted into decibel. These metrics are evaluated using the values presented in Table \ref{tb:setup}. 

We evaluate the molecular output signal throughput (see Figure \ref{fig:throughput}) for $72$ hours (this is sufficient time to produce detectable levels of $[sfGFP]$),  bacterial population of $10,000$ cells, a diffusion coefficient of $2\times10^{-7}$ (similar value to the model investigated in \cite{Kubitscheck2000}), and a range of initial concentrations for the $[sfGFP]$ from $10^{-2}$ to $10^{-6}$ mol/L. It can be noted for this scenario, that for the first $10$ hours there is a steep increase in the production of the molecular output signal and a flattening in the throughput after $20$ hours for most of the curves. The exception is for the $10^{-2}$ curve, where this saturation process only occurs after $50$ hours as the positive feedback loop requires more time to balance all the chemical reactions related to this process due to the high initial value of $[sfGFP]$.

\begin{figure}[t!]
  \includegraphics[width=\columnwidth]{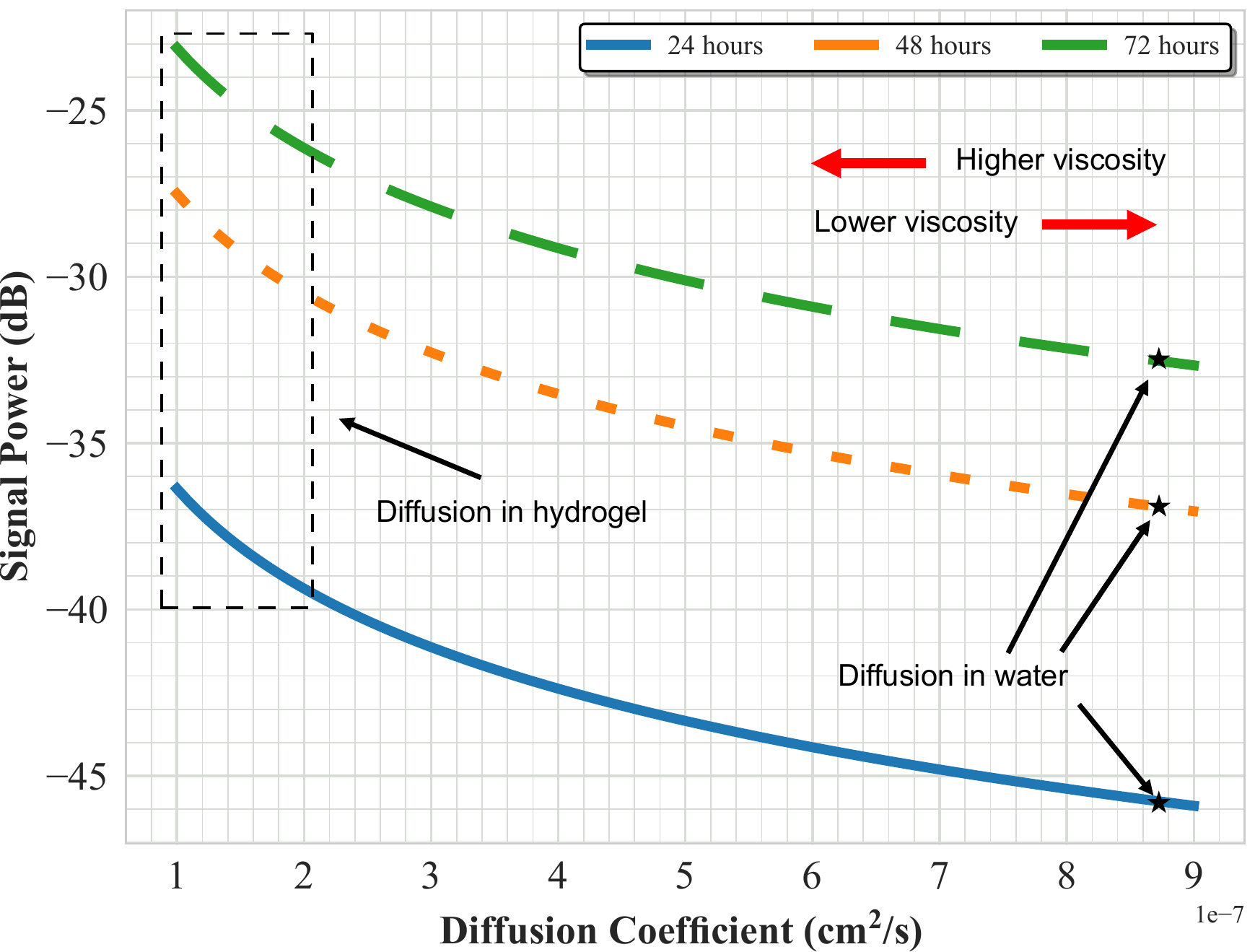}
  \caption{Evaluation of the molecular output signal power that reaches the hydrogel edge membrane for low and high viscosity medium.}
  \label{fig:power}
\end{figure}

We modified the scenario for the throughput analysis, to investigate the signal power that reached the edge membrane of the hydrogel. In this case, we considered the hydrogel as a bubble with diameter of $7$ mm, an initial concentration for the molecular output signal $[sfGFP]$ of $10^{-6}$ mol/L; a range of diffusion coefficient values from $1\times10^{-7}\,\text{cm}^2/\text{s}$ to $2\times10^{-7}\,\text{cm}^2/\text{s}$, and evaluated using (\ref{eqn:diffusion}), where we then converted into $db$ (assuming a 1:1 ratio between mol/L and watts), at $24$, $48$ and $72$ hours. In Figure \ref{fig:power} we can see that the signal power is higher for lower diffusion coefficients (higher viscosity), and in particular for hydrogels compared to  media with lower viscosity, such as water. Therefore, this result infers that the hydrogels or other higher viscosity material can be used to possibly direct molecular signals produced by the bio-nanomachine transmitters, while the lower viscosity media can be applied for omnidirectional propagation of molecular signals.

\begin{table}[t!]
  \centering
  \caption{Parameters used to evaluate Equations (\ref{eqn:dA})-(\ref{eqn:diffusion})}
  \label{tb:setup}
  \begin{tabular}{| c | c | c |}
    \hline
    Variable & Value & Unit\\ \hline
    \hline
    $c_A$ & $2.7\times10^{-2}$ & $\text{mmol/L}$\\
    $c_R$ & $2.7\times10^{-2}$  & $\text{mmol/L}$\\
    $k_A$ & $2\times10^{-3}$  & $\text{d}^{-1}$\\ 
    $k_R$ & $2\times10^{-3}$  & $\text{d}^{-1}$\\ 
    $k_0$ & $1\times10^{-2}$ & $\text{d}^{-1}$\\
    $k_1$ & $0.1$ & $\text{d}^{-1}$\\
    $k_2$ & $0.1$ & $\text{d}^{-1}$\\ 
    $k_3$ & $1\times10^{-2}$ & $\text{d}^{-1}$\\ 
    $k_4$ & $0.1$ & $\text{d}^{-1}$\\ 
    $k_5$ & $0.1$ & $\text{d}^{-1}$\\\
    $k_{sfGFP}$ & $1\times10^{-3}$  &$\text{d}^{-1}$\\ 
    $K_A$ & $2\times10^{-3}$ & $\text{gm}^{-3}$\\ 
    $K_R$ & $2\times10^{-3}$ & $\text{gm}^{-3}$\\ 
    $K_C$ & $1$ & $\text{gm}^{-3}$\\ 
    $K_N$ & $1$ & $\text{gm}^{-3}$\\ 
    $\mu_{1}$ & $1\times10^{-4}$ & $\text{gm}^{-3}$\\   
    $m_1$ & $1\times10^{-4}$ &$\text{gm}^{-3}$\\ 
    $U_{AHL}$ & $2\times10^{-2}$ &--\\
    $U_{G1}$ & 0.6 &--\\
    $U_{LuxR}$ & $2\times10^{-2}$ &--\\
    $U_{sfGFP}$ & $5\times10^{-2}$  &--\\ 
    $D_h$ & from $1\times10^{-7}$ to $9\times10^{-7}$ &$\text{cm}^{2}/\text{s}$ \\
    $t_h$ & $72$ & hours \\
    $x_h$ & $7\times10^{-3}$ &$\text{m}$ \\\hline
    \end{tabular}
\end{table}

\section{Experimental Analysis}\label{sec:exp_analysis}

\begin{figure}[t!]
  \includegraphics[width=\columnwidth]{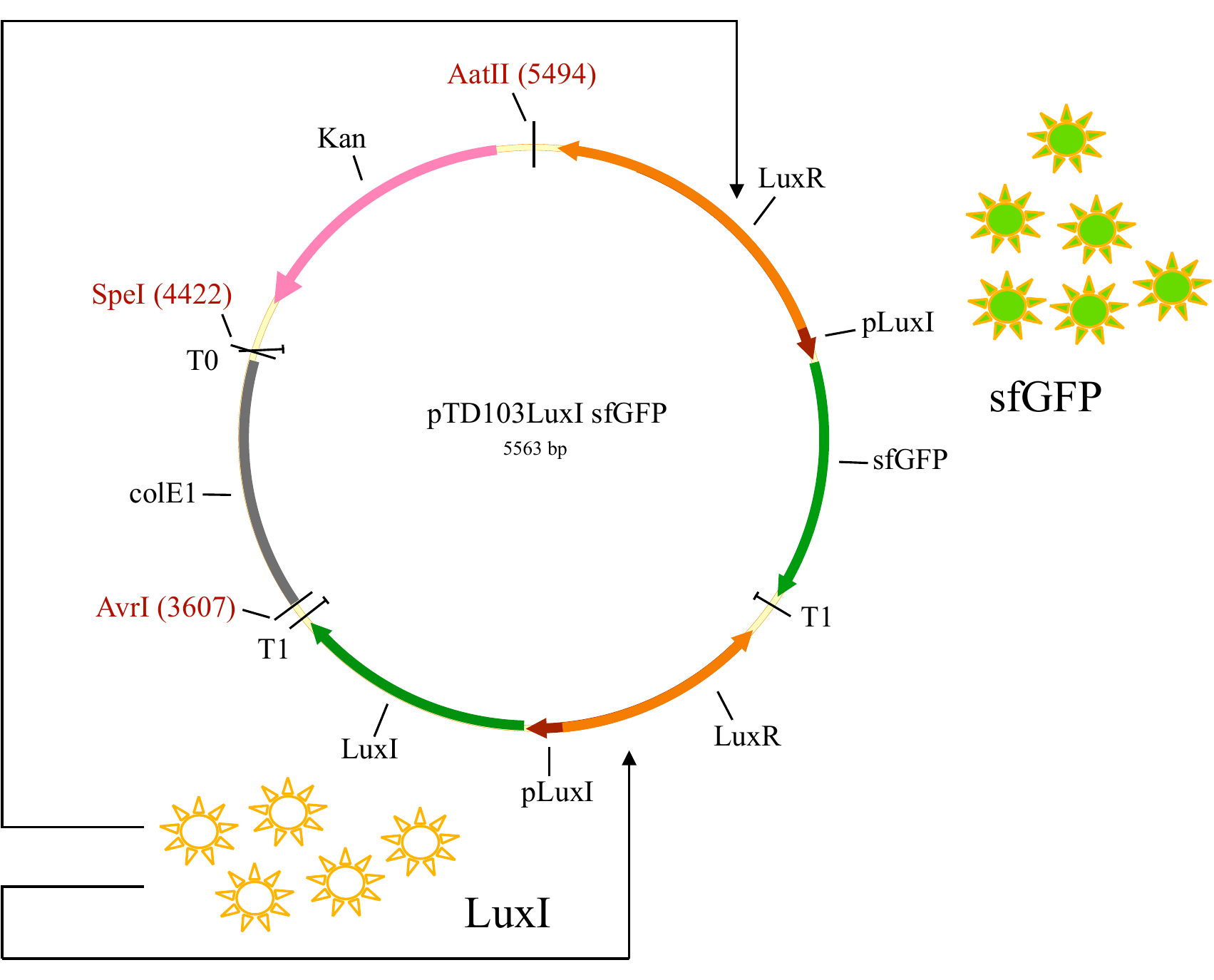}
  \caption{Plasmid structure used in this study to produce sfGFP due to the increase in production of the molecular input signal LuxI. Adapted from \cite{Prindle2012}.}
  \label{fig:gene_bio}
\end{figure}

\new{We prepared wetlab experiments to observe the operation of the positive feedback oscillator in terms of the amount of molecular output signal produced $[sfGFP]$. Bacteria used in this experiment was \textit{E.coli} DH5$\alpha$ pTD103luxI\_sfGFP (abbreviated as \textit{E. coli} pTD103). \new{We purchased Luria-Bertani (LB)} broth from Fisher Bioreagents\textsuperscript{\texttrademark)}, Agarose from Promega\textsuperscript{\texttrademark}, and Calcium chloride hexahydrate from Merck \textsuperscript{\texttrademark}. Other reagents and antibiotics required for this experiment were purchased from Sigma-Aldrich\textsuperscript{\texttrademark}. Kanamycin (Kan) and Chloramphenicol (Cam) were sterile and utilised at concentrations of $50\,\mu \text{g/ml}$ and $34\,\mu \text{g/ml}$, respectively. The plasmid used in this experiment, pTD103luxI\_sfGFP, was a gift from Jeff Hasty (Addgene plasmid $\#48885$; http://n2t.net/addgene:48885;RRID: Addgene\_48885), see Figure \ref{fig:gene_bio} \cite{Prindle2012}.}

\begin{figure*}[t!]
  \includegraphics[width=0.8\textwidth]{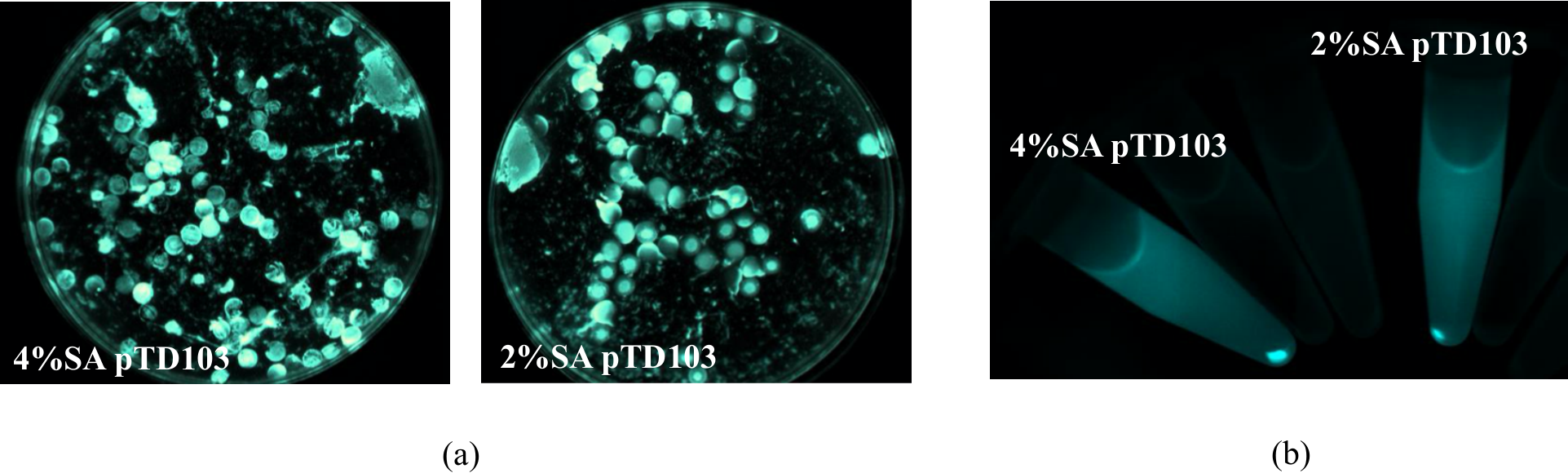}
  \caption{Transmission of the molecular signal into the liquid medium after 72 hrs incubation. Cells were captured in 4\% and 2\% SA hydrogel, then incubated with LB-Kan broth for 72 hrs without shaking. The experiment was performed once. The images were taken by iBright FL1000 system (Invitrogen). (a) Observation of the spatial distribution of the molecular output signal in a plate. (b) Observation of the molecular output signal production in test tubes, where the engineered bacterial population is placed at the bottom of the tubes.}
  \label{fig:tx_res}
\end{figure*}

\begin{figure}[t!]
  \includegraphics[width=\columnwidth]{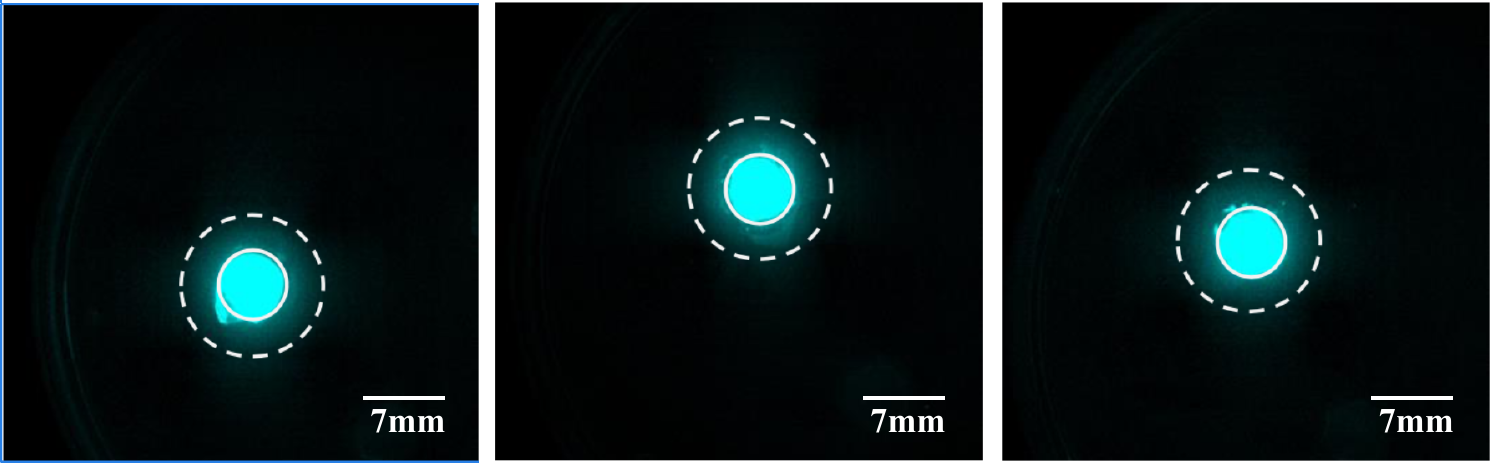}
  \caption{Secreted molecular output signals recorded after $72$ hours. The cells were encapsulated in agarose hydrogel bubble and then incubated with $1.5\%$ agarose-Cam for $72$ hours. The experiment was performed as triplicates, and the images were taken by iBright FL1000\textsuperscript{\texttrademark} system (Invitrogen\textsuperscript{\texttrademark}).}
  \label{fig:expbio_res}
\end{figure}

\new{In our first experiment, we encapsulated \textit{E. coli} pTD103 in sodium alginate (SA) hydrogel and detected fluorescent proteins secreted into the liquid medium. \textit{E. coli} pTD103 stock \new{was} streaked onto LB-Kan agar and incubated at $37^{\circ}\text{C}$ $24-48$ hours as a fresh growth then cultivated into LB-Kan broth and incubated at $37^{\circ}\text{C}$ for $4$ hours at $37^{\circ}\text{C}$, $220$ rpm in a benchtop orbital shaker incubator (Grant-bio\textsuperscript{\texttrademark}) prior to a centrifugation at $1,500$ rpm for $10$ minutes (Sigma 4K15\textsuperscript{\texttrademark}, mid bench centrifuge). Cell pellets were harvested, re-suspended in $2\%$ and $4\%$ SA solution and dropped into $50\,\text{mM}\,\text{CaCl}^2$ solution to form SA beads. The beads were stored at $4^{\circ}\text{C}$ overnight prior to the setting up of the experiments.}

\new{On the day of experiment, the beads were washed twice by sterile distilled water then incubated in LB-Kan broth at $37^{\circ}\text{C}$ for $72$ hours without shaking. Supernatants were harvested and centrifuged at $20,000$ times the gravity force for $20$ minutes at $4^{\circ}\text{C}$ to remove cell pellets and other residues. This procedure ensures that the detected molecular output signal was the one propagated into the media, not inside the cells or any SA residues.
The molecular output signal was detected and recorded by iBright FL1000 system \textsuperscript{\texttrademark} (see Figure \ref{fig:tx_res}). In this experiment, we utilised SA hydrogel due to its permeability of nutrient infusion for cell growth, as well as its usage as a protective layer from environmental hazards  \cite{Anselmo2016,Nicodemus2008}. We used two concentrations of SA, $2\%$ and $4\%$, to observe the cell growth and production of the molecular output signal within the SA beads and into the liquid medium. As it can be noted from Figure \ref{fig:tx_res}a, the molecular output signal propagates with better performance at $4\%$ SA in comparison to the $2\%$ SA. A similar result is shown in Figure \ref{fig:tx_res}b, where the molecular output signal is shown in the solid and in the supernatant parts at the bottom and at the top of centrifuge tubes, respectively. This suggested that cells secreted fluorescent proteins while being encapsulated within the SA hydrogel beads. This phenomenon was also observed in other studies \cite{Li2017encapsulation,Tang2020}. Nevertheless, we cannot deny a possibility that the cells may have escaped from the hydrogel beads and produced fluorescent proteins while growing in medium.} 

\new{To confirm that E.coli pTD103 can secrete fluorescent proteins  [\emph{sfGFP}] into their surrounding environment without escaping, we performed another experiment with solid medium (i.e., agarose hydrogel). For this experiment, the \textit{E.coli} pTD103 were cultivated in LB-Kan broth for $3-3.5$ hours at $37^{\circ}\text{C}$, $220$ rpm. The cell pellets were harvested by  centrifugation at $1,500$ rpm for $5$ min (Sigma 4K15\textsuperscript{\texttrademark}, mid bench centrifuge), re-suspended then mixed well in warm $1.5\%$ agarose-Kan solution prior to pouring onto $\phi100$ mm petri dishes  (Sarstedt\textsuperscript{\texttrademark}). Once the \new{bacteria-agarose mixture} solidified, the petri dishes were incubated at $37^{\circ}\text{C}$ for $24$ hours \new{subsequent steps}. Each bacteria-hydrogel bubble was placed onto a petri dish containing 1.5\% agarose-Cam and incubated at $37^{\circ}\text{C}$ for $24$, $48$ and $72$ hours. At each time point, the fluorescent signals were recorded by iBright FL1000 system\textsuperscript{\texttrademark} (Invitrogen\textsuperscript{\texttrademark}). At the $72$ hours time point, a triplicate was performed, and the result can be seen in Figure \ref{fig:expbio_res}. The bacteria were entrapped within $1.5\%$ agarose gel containing nutrient (LB) as well as antibiotics (Kan) in order to maintain cell growth as well as their ability to  produce the desired proteins  $[sfGFP]$). According to our results, after $24$ hours incubation, the bacteria grew well and started to produce the molecular output signal (data not shown). Once  placed into a solid medium containing $1.5\%$ agarose-Cam, after $72$ hours, a ring of fluorescence was observed at the edge of each hydrogel bubble, suggesting the cells can grow and produce molecular output signal which are diffused out of their agarose hydrogel bubble (Figure \ref{fig:expbio_res}). The experiment was performed with triplicates to ensure its consistency. We eliminated the ability for the bacteria to mobilise out of their bubbles by using Cam, an antibiotic preventing \textit{E. coli} pTD103 growth (data not shown).}

\section{Conclusions}\label{sec:conclusions}

\new{The internal machinery of bacteria} have been engineered with the purpose of designing biocompatible technological applications. In this paper, we investigated the engineering of \new{a} bacterial quorum sensing system to create a positive feedback loop required for the design of a bio-nanomachine transmitter. In addition, we propose to encapsulate the cells in hydrogel to protect them from mixing with the natural cells in the environment, and this solution can pave the way for  intrabody molecular communications systems using bacterial signalling in humans and animals. 

Our numerical analysis shows, considering the scenario investigated, a high throughput with respect to time (reaching more than $10^{-2}\,\text{mol/L}$ in $72$ hours). It also highlights the saturation of the molecular output signal production after $20$ hours for most of the cases, which can be applied for future biotechnological systems that require this stability phase of the bio-nanomachine transmitter operation. When observing the relationship between the signal power at the hydrogel edge membrane and the different values of molecular diffusion coefficients (i.e., higher or lower viscosity media), it can be inferred that the hydrogels can facilitate directive propagation, resulting in higher molecular output signal power than the conventional free-diffusion propagation.

We also provided a proof-of-concept of the bio-nanomachine transmitter through wet lab experiments, where the bacterial population is shown to produce higher fluorescence in their close vicinity, and is able to propagate the molecular output signal through the hydrogel medium and into the environment. It is our intention to further expand the design introduced here to create an end-to-end molecular communications systems that can be used for biosensing and biocomputing applications. 

\begin{acks}
This publication has emanated from research conducted with the financial support of Science Foundation Ireland (SFI) and the Department of Agriculture, Food and Marine on behalf of the Government of Ireland under Grant Number [16/RC/3835] - VistaMilk.  
\end{acks}

\bibliographystyle{ACM-Reference-Format}
\bibliography{nanococoa}


\end{document}